\documentstyle[aaspp,11pt]{article}

\lefthead{Armitage \& Pringle}
\righthead{Warping of protostellar accretion disks}

\begin{document}

\title{RADIATION INDUCED WARPING OF PROTOSTELLAR ACCRETION DISKS}

\author{P. J. Armitage}
\affil{Canadian Institute for Theoretical Astrophysics, McLennan Labs,
	60 St George St, Toronto, M5S 3H8, Canada}
\and
\author{J. E. Pringle}
\affil{Institute of Astronomy, Madingley Road, Cambridge, CB3 0HA, UK}

\begin{abstract}
We examine the consequences of radiatively driven warping of
accretion disks surrounding pre-main-sequence stars. These disks
are stable against warping if the luminosity arises from a steady
accretion flow, but are unstable at late times when the intrinsic
luminosity of the star overwhelms that provided by the disk. 
Warps can be excited for stars with luminosities of around $10 L_\odot$ 
or greater, with larger and more severe warps in the more luminous
systems. A twisted inner disk may lead to high extinction towards 
stars often viewed through their
disks. After the disk at all radii becomes optically 
thin, the warp decays gradually on the local viscous timescale, 
which is likely to be long. We suggest that radiation induced warping 
may account for the origin of the warped dust disk seen in Beta
Pictoris, if the star is only $\sim 10-20$ Myr old, and could
lead to non-coplanar planetary systems around higher mass
stars.

\medskip
\noindent{\em ApJ Letters, in press}

\end{abstract}	


\newpage

\section{INTRODUCTION}
A geometrically thin, optically thick accretion disk is unstable to
self-induced warping when illuminated by a sufficiently strong central radiation
source (Pringle 1996; Maloney, Begelman \& Pringle 1996; see also
Petterson 1977). For a steady accretion disk, generating luminosity
with an efficiency $\epsilon = L / \dot{M} c^2$, the instability
can readily be shown to be important for disks around neutron stars and 
black holes, where it is generally assumed that $\epsilon \sim {\cal O} (0.1)$,
provided only that the inner regions of such disks are not advectively
dominated. Previous authors have discussed the implications of the
instability for X-ray binaries, the masing disk in NGC4258 
(Maloney, Begelman \& Pringle 1996; Maloney \& Begelman 1997),
and the Unified Scheme for Active Galactic Nuclei (Pringle 1997). 
For disks around less compact objects, where $\epsilon$ is orders
of magnitude smaller, steady disks are predicted to be stable.

The conclusion that the warping instability is unimportant for disks 
around white dwarfs, or ordinary stars, can be evaded if
the luminosity of the central source greatly exceeds that 
provided by accretion. The disk then sees a radiation source
much more powerful than that implied by the local accretion rate,
and can be warped in the same way as a disk around a more compact
body. Even if the luminosity {\em is} generated solely by accretion,
a mass flux that is a strongly decreasing function of radius
could allow the outer parts of a disk to be warped by the stronger
radiation emitted from the inner disk. Such a scenario might
be important in outbursting systems such as dwarf novae and
FU Orionis stars, where the disk may be subject to thermal instability.

In this Letter, we apply the warping instability
to accretion disks surrounding pre-main-sequence stars. In \S 2 we
show that these disks are stable against warping during the main
phase of steady disk accretion, but become unstable
at late times when the surface density 
and accretion rate fall to low values. In \S 3 we apply
our results to $\beta$ Pictoris, 
which has been observed to be surrounded by a warped dust disk
(Burrows et al. 1995),
and show that this warp is consistent with a radiative origin.
In \S 4 we present a numerical calculation of the excitation and
subsequent decay of the warp. Our conclusions are summarized
in \S 5.

\section{WARPS IN PROTOSTELLAR DISKS}
\subsection{Conditions for the warping instability}
The conditions under which a thin accretion disk, optically thick
to absorption and re-emission of radiation, is unstable to radiation
induced warping are derived in Pringle (1996) and Maloney,
Begelman \& Pringle (1996). Defining the timescales for the viscous 
and radiation torques as,
\begin{equation}
 t_{\nu_2} = {{2 R^2} \over {\nu_2}}, \ \ \ \ 
 t_\Gamma = {{12 \pi \Sigma R^3 \Omega c} \over {L_*}},
\label{eq2}
\end{equation} 
instability occurs when the ratio, 
\begin{equation}
 {{t_\Gamma} \over {t_{\nu_2}}} < \gamma_{\rm crit}.
\label{eq3}
\end{equation} 
Here $L_*$ is the luminosity of the central source, $\Sigma$ the
disk surface density, $\Omega$ the Keplerian angular velocity,
and $\nu_2$ the $(R,z)$ component of the disk viscosity. We note that $\nu_2$
need not be equal to the more familiar $(R,\phi)$ component
of the viscosity $\nu_1$ (e.g. Pringle 1992), and we denote
the ratio as $\eta=\nu_2 / \nu_1$. In an earlier paper (Pringle 1996) 
$\gamma_{\rm crit}$ was estimated to be $\sim {1 / 2 \pi}$,
and this is consistent with the numerical calculations in \S 4.  

For radii much greater than the stellar radius, $R_*$,
$\nu_1 \Sigma = \dot{M} / (3 \pi)$. From equation (\ref{eq3}), the luminosity
required to warp the disk at radius $R$ is then,
\begin{equation}
 L_* \gtrsim {{2 \dot{M} \eta R \Omega c} \over {\gamma_{\rm crit}}}.
\label{eq4}
\end{equation} 
If this luminosity is provided by the accretion disk and boundary
layer, then $L_* \simeq G M_* \dot{M} / R_*$, and for a solar mass star the
critical radius beyond which the disk will be unstable to
warping is,
\begin{equation}
 R \gtrsim 3.5 \times 10^5  \  \eta^2 
 \left( {R_* \over R_\odot} \right)^2 \ {\rm a.u.} \ \gg R_{\rm disk},
\label{eq6}
\end{equation}
taking $\gamma_{\rm crit} = 1/2 \pi$.
Steady protostellar disks are therefore, as expected, stable
against warping induced by their own luminosity.

If the luminosity is {\em not} provided by the accretion flow,
equation (\ref{eq4}) implies,
\begin{equation}
 R \gtrsim 3 \ \eta^2 \left( {M_* \over M_\odot} \right)
 \left( { \dot{M} \over {10^{-9} \ M_\odot {\rm yr}^{-1}} } \right)^2
 \left( {L_* \over {10 \ L_\odot}} \right)^{-2} \ {\rm a.u.}
\label{eq7}
\end{equation}
Typical accretion rates in pre-main-sequence T Tauri stars 
are $\sim 10^{-7}-10^{-8} \ M_\odot {\rm yr}^{-1}$, so with
$L_* \sim 10 L_\odot$ this implies that $\dot{M}$ needs to fall by 
1-2 orders of magnitude before the inner parts of the disk become
potentially unstable to warping. A large value of $\eta$ at a few
a.u. would suppress the instability.
Whether warping occurs will then
depend on the growth rate of the instability 
and, critically, on whether the disk remains
optically thick to re-emission of intercepted stellar radiation.
At radii where the disk is optically thin the re-emitted flux will
be isotropic, so that there is no radiation torque or possibility
for a warp to develop. At large radii of $\sim 10^2$ a.u. and greater,
protostellar disks are known to be optically thin in the mm
emission characteristic of those radii, but the inner few a.u. 
are likely to be extremely optically thick. Models for FU Orionis
outbursts, for example, suggest that typical disks around solar
type stars have a Shakura-Sunyaev viscosity parameter $\alpha \ll 1$
in the very innermost regions (Bell \& Lin 1994), with correspondingly
high optical depths. At radii where
dust remains present, the implied surface densities suggest that the disks
will become unstable to warping in their inner regions
before they become optically thin. 

For more massive Herbig Ae-Be stars, with
higher luminosities, the critical accretion rate below which warping
occurs will be greater. We would thus not
{\em expect} to observe flat passive (reprocessing) disks around stars
of $M_* \gtrsim 2 \ M_\odot$, as these would be destroyed by
the action of the twisting instability. This could lead to a substantial
fraction of systems where the star was viewed {\em through}
the warped disk, displaying high extinction at short wavelengths.
 
\subsection{Growth and decay timescales}
The growth rate of the instability is $\sim t_\Gamma$ (Pringle 1996).
Equation (\ref{eq2}) yields an estimate for the growth time,
\begin{equation}
 t_\Gamma \simeq 4 \times 10^3 \ 
 \left( {M_* \over {2 M_\odot}} \right)^{1/2}
 \left( {R \over {3 \ {\rm a.u.}}} \right)^{3/2}
 \left( {L_* \over {10 \ L_\odot}} \right)^{-1}
 \left( {\Sigma \over {1 \ {\rm gcm}^{-2}}} \right) \ {\rm yr}.
\label{eq8}
\end{equation} 
A surface density in gas,
$\Sigma=1 \ {\rm gcm}^{-2}$, corresponds roughly to $\tau \sim 1$
using a typical dust opacity at 100 $\mu$m (Men\' shchikov \& 
Henning 1997), and a standard gas to dust ratio of 100.
This timescale is evidently much shorter than both the typical
lifetime of a pre-main-sequence disk, which is a few Myr (Strom
1995), and the estimated disk dissipation timescale of $\sim 10^5$ 
years (Wolk \& Walter 1996). 

Equation (\ref{eq8}), together with the instability criterion, equation (\ref{eq7}),
specify the qualitative development of warping in protostellar disks. Radii
smaller than the critical radius, or so large that the disk is always
optically thin to re-emission of stellar radiation, are stable. At all 
unstable radii, the strongest warping is expected when
the surface density is lowest, just before the disk becomes optically
thin. How far the warp is able to develop then depends on the time
available before the annulus becomes optically thin, which to
order of magnitude will just be the viscous time {\em of the disk}.
The {\em local} viscous time will be shorter,  
but it is the viscous time of the entire disk
that is relevant as this is the timescale on which $\dot{M}$, and
hence also $\Sigma$ decays. The strongest twisting will
occur in the inner disk, both because $t_\Gamma$ is
smallest for the unstable modes there, and because inflow
through the disk can advect an outer warp to smaller radii.

After a given annulus in the disk has become optically thin, 
radiative forcing of the warp ceases. The twist will then
decay on the viscous timescale of the disk, spreading
as it does to larger radii than the 
originally unstable region. If the original twist is severe,
many viscous times may be required to flatten the inner and
outer parts of the disk into the same plane. The
decay timescale itself is likely to be long, as the viscous timescale  
is already $\sim$ 1 Myr when there is still a significant 
amount of gas present, and thus the disk could retain a modest
warp for a lengthy period after radiative excitation of the warp
had ceased.

\newpage
\section{ESTIMATES FOR A $\beta$ PICTORIS PROGENITOR}

In this Section we consider $\beta$
Pictoris, which is observed to be surrounded by a dust disk. {\em 
Hubble Space Telescope} observations show that the
inner (few 10's of a.u.) region of the disk is warped by several
degrees with respect to the outer disk (Burrows et al. 1995), and
we examine whether a warp of this extent could arise from a
radiatively driven mechanism.

For the stellar parameters, we follow
Lanz, Heap \& Hubeny (1995), and take a mass $M_* = 1.8\ M_\odot$,
an effective temperature $T_* = 8200 \ {\rm K}$, and a bolometric
luminosity $L_* \approx 11.3 \ L_\odot$. $R_*$ is then $\approx 1.7 \ R_\odot$.
We assume that the condition for an annulus in the
disk to be sufficiently optically
thick to re-emission is that $\tau = 1$ at the peak 
of the $\nu F_\nu$
spectral energy distribution at the temperature, $T_{\rm disk}$, of
that annulus. $T_{\rm disk}$ is estimated 
using the temperature distribution for a flat
reprocessing disk given by Kenyon \& Hartmann (1987). For the
stellar parameters of $\beta$ Pictoris, the wavelength of
maximum disk emission, $\lambda_{\rm max}$, is in the 40-200 $\mu$m
region for disk radii between 1 and 10 a.u. 

The maximum susceptibility to warping occurs
when the disk is only just optically thick.
The surface density for this can be estimated from
$\lambda_{\rm max}$ once the opacity is specified. We consider
three forms for the dust opacity; those used by Rowan-Robinson (1986),
Ossenkopf (1993), and Men\' shchikov \& Henning (1997, who show
all three in their Fig.~15). Using these functions, kindly provided
to us by Dr Men\' shchikov, the critical surface density in
dust, $\Sigma_{\rm crit}$, for the disk to be optically thick is 
shown as a function of
radius in Fig.~1. There is reasonable agreement between the
various opacities in the region of interest.

In addition to the uncertain opacity, 
there is a larger uncertainty arising from the
possible changes in the dust properties within the disk. For
example, agglomeration of the dust into larger particles
could drastically reduce the opacity. We crudely allow for
this by keeping the gas to dust ratio, $f_{\rm g}$, as a free
parameter, so that the total surface density at the optically
thick to thin transition is $\Sigma_{\tau =1} = f_{\rm g} \Sigma_{\rm crit}$. 
A large $f_{\rm g}$ then implies a depleted dust opacity.

The radial extent of radiatively induced warping may then be
estimated as follows. We assume that there is some timescale,
$t_{\rm clear}$, available for the development of a warp as 
the disk becomes optically thin,
and that the outer optically thick annulus of the disk is
unstable to warping by equation (\ref{eq7}) as this transition
is reached.
Requiring that $t_\Gamma < t_{\rm clear}$ for $\Sigma = \Sigma_{\tau=1}$
then fixes a maximum radius, $R_{\rm warp}$, for a given 
$f_{\rm g}$ and $t_{\rm clear}$.
This is shown for $t_{\rm clear} = 10^5$ and $10^6$ yr in Fig.~1, again
for the three opacity functions described above. As expected, there
is a fairly strong dependence on $f_{\rm g}$ - a high gas to dust
ratio severely restricts the radial extent of possible warping. 
However, for a clearing timescale of 0.1-1 Myr, and $f_{\rm g} \sim 10^2$, we
find that a warp of 10-40 a.u. extent should develop, consistent
with it being the progenitor of the observed warped disk around 
$\beta$ Pictoris.

\newpage
\section{NON-LINEAR DEVELOPMENT OF THE WARP}

To verify these estimates we have computed the evolution
of the warp numerically. The code used is identical to that described
by Pringle (1997), except for the inclusion of a variable radius
beyond which the disk is optically thin. This radius is computed
from the surface density using the form for $\Sigma_{\rm crit}$
shown in Fig. 1, and assuming $f_{\rm g} = 10^2$. The
viscosity is taken as $\nu_1 = \nu_2 = \nu_{10} R^{3/2}$, 
implying a steady-state surface density profile, $\Sigma \propto
R^{-3/2}$, of the form adopted in `minimum-mass' solar nebula
models (Weidenschilling 1977, Hayashi 1981, for a review see Lissauer 
1993). Taking
$\nu_{10} = 1.5 \times 10^{-6} \ {\rm cm}^{1/2} {\rm s}^{-1}$
ensures a disk mass within 250 a.u. of $\sim 0.1 \ M_\odot$
and an evolutionary timescale of $\sim 1$ Myr, consistent
with observations of T Tauri stars (Beckwith et al. 1990; Strom 1995). 
We use 150 radial and 120 azimuthal grid points, with the radial mesh
logarithmically spaced such that the outer edge is at 250 a.u.
The stellar mass and luminosity are taken from Lanz, Heap \&
Hubeny (1995). The disc is initially flat except for a small,
seed warp of less than one degree in the outer regions, with a
steady-state accretion rate of $10^{-7} \ M_\odot {\rm yr}^{-1}$.
This disk is then allowed to drain freely onto the star.

Fig.~2 and Fig.~3 summarize the results. After around 7 Myr,
a warp is excited at an initial radius of around 5 a.u. 
The growth is rapid on the scale of the Figure, as the
growth timescale at this radius is much less than the viscous
timescale at the outer edge. The warp steadily diffuses outwards, 
with a complicated pattern of growth occurring in the 
inner regions. The excitation of the warp 
lasts for a few Myr, during which time a large peak tilt
develops in the inner 10 a.u. As was found in the
calculations presented by Pringle (1997) the
disk tilt remains a smooth function of radius throughout -- i.e
the disk does not become so twisted as to lose its integrity.
Tests at a variety of numerical resolutions show variations
in the disk evolution in the strongly non-linear regime, this
calculation shows a typical result but cannot be regarded 
as a prediction of, say, the maximum inclination attained.
The large uncertainties in $\nu (R)$, $\eta (R)$, $\tau (R)$, and the 
complicating effects of a disk with significant scale height,
reinforce that conclusion.

Once the disk has become optically thin, the warp decays on
the viscous timescale.
Fig.~3 shows the profile of the warp after $\sim 6$ Myr
of decay, when the warp of the inner few 10's of 
a.u. is $\sim 5-10$ degrees. At this stage the disk is
not tightly `wound-up', and is decaying with a simple warp.
The inner regions for which the
viscous timescale is shorter have already flattened themselves
out, and the warp decays smoothly towards the outer disk 
edge. For this surface density profile, with $\Sigma \propto R^{-3/2}$,
the relatively small reservoir of mass at large radii is
unable to ultimately flatten the disk back into its original
plane. At the end of the calculation, the disk is
thus close to flat but tilted with respect 
to its original plane.

\newpage
\section{DISCUSSION}

In this Letter, we have discussed the possibilities for radiation
driven warping of protostellar disks. We find that as
a result of the high optical depth of the inner disk at typical
accretion rates, the disk should be unstable to warping at late
times, when the accretion luminosity is negligibly small compared
to that of the star. The warp is excited in the optically
thick inner disk, and is most severe there, 
but spreads to a range of radii. 

The development of a warp is a strong function of the stellar luminosity,
and hence mass. As the luminosity increases, warping should set in at 
progressively higher accretion rates. Twisted inner disks may
be a partial cause of the lack of the usual 
accretion disk signatures in Herbig Ae-Be stars, and
could lead to anomalously high extinction towards stars
that are actually several Myr old. 

For the Sun, the protosolar luminosity is expected to be too
low for warping to occur before planet formation was well 
underway. This is consistent with
the approximate coplanarity of planetary orbits. Around more
massive stars, of several solar masses, the disk lifetime
is shorter, and the higher luminosity means that the instability 
occurs at higher accretion rates. This mechanism could then lead 
to non-coplanar planetary systems around massive stars.

Applying this model to a $\beta$ Pictoris type star, we find that the 
disk should have become warped at the
time when the inner disk reached the optically thick 
to thin transition, and that the radial extent of the predicted warp
is a few 10's of a.u. Numerical calculations
confirm these estimates, and suggest that the warp at its peak
might have been severe. Moreover,
the final plane of the disk may not coincide with that of
the stellar equator. In this scenario, the current small warp 
seen in $\beta$ Pictoris could be the decaying remnant of a 
previously much more severe twist induced by the radiative instability.

Since the prevailing explanation for the warp in the $\beta$
Pictoris disk appeals to a large planet on an inclined orbit
(Burrows et al. 1995), it is worth commenting on the 
differences between this scenario and that presented in \S3. 
Our work suggests that warps should occur in dusty disks
around luminous stars; whether it can be applied to the
specific case of $\beta$ Pictoris depends on the star's age.
If the system is of the order of $10^8$ years old (Burrows et al. 1995; 
Brunini \& Benvenuto 1996), then a planet is a plausible
mechanism to maintain the observed warp. Conversely, if $\beta$ Pictoris
is only $\sim$ 12 Myr old, as inferred by Lanz, Heap \& 
Hubeny (1995), then no planet is required to explain the current
warped disk. Weak support for a young age is provided by 
observations of the analog system HR 4796A, which is
inferred to be $\sim 8$ Myr old (Stauffer et al. 1995). 

This model predicts that warps should be
common in the inner regions of protostellar disks at late epochs,
with the largest and strongest warps occurring in disks
around high luminosity stars. Warping can only be
avoided if either the disk becomes optically thin to re-emission 
at higher accretion rates than is currently favored, or the
dust opacity at late times is severely depleted. Although few
systems may be as fortuitously aligned for observations as the disk
in $\beta$ Pictoris, the changes in the scattering properties
implied by warped disks may be detectable in a larger
sample of young stars.

\begin{figure}[tb]
 \plotone{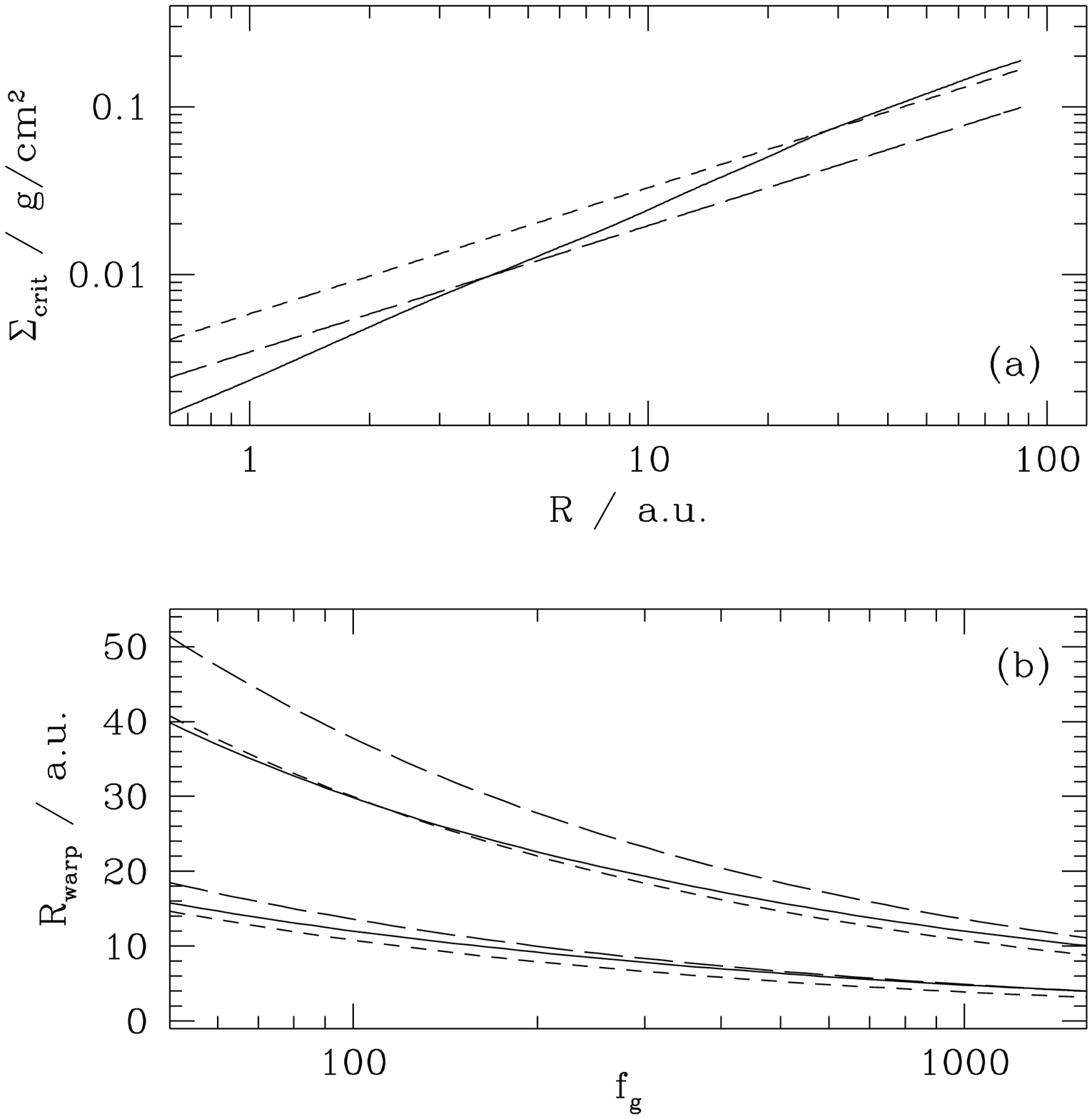}
 \caption{Properties of a disk surrounding a $\beta$ Pictoris-type
 star at an earlier stage of its evolution. (a)
 The critical surface density {\em in dust} required for the disk
 to be optically thick to re-emission of stellar radiation, for
 three dust opacities described in the text. (b)
 The radial extent of the warp, assuming that the time
 available for the warp to grow as the disk becomes optically thin
 is $10^5$ yr (lower curves), or $10^6$ yr (upper curves), as a
 function of the gas to dust ratio $f_{\rm g}$.} 
\end{figure}

\begin{figure}[tb]
 \plotone{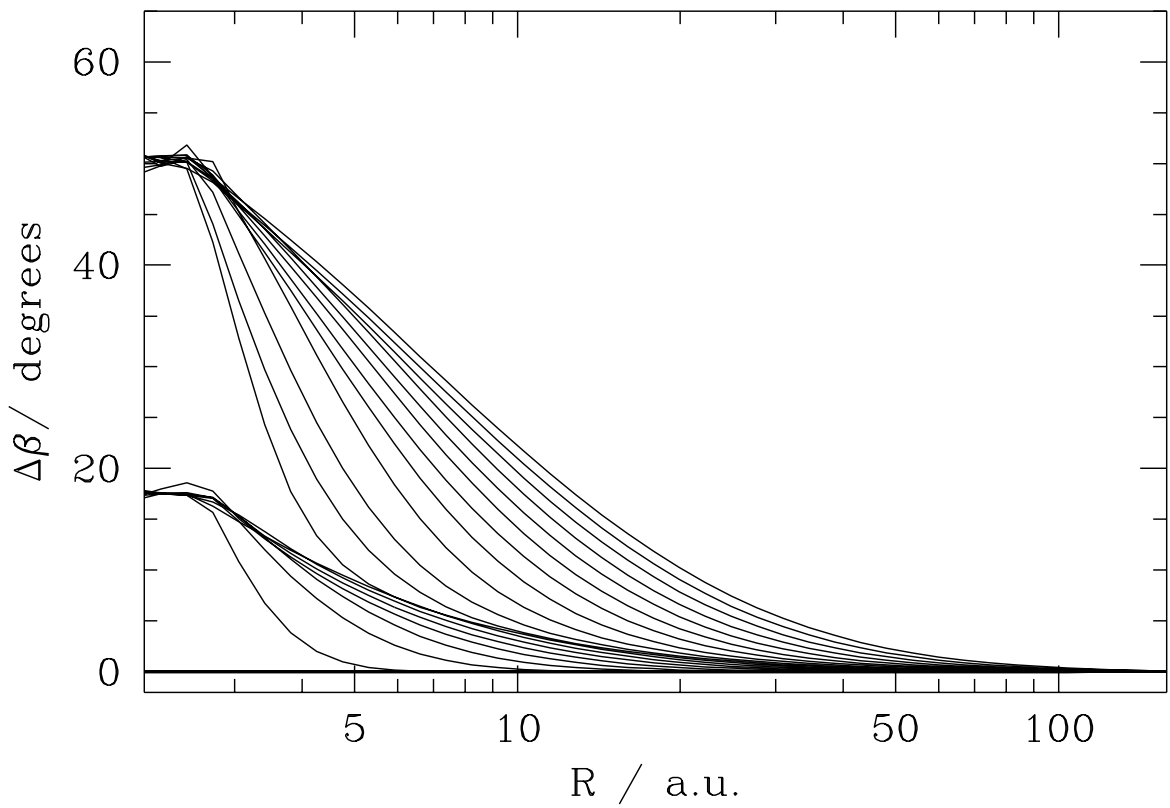}
 \caption{The tilt of annuli in the disk, relative to the outermost
 annulus at 250 a.u., during the initial growing phase of the instability.
 Curves are plotted at 0.5 Myr intervals, for 2Myr.}
\end{figure}

\begin{figure}[tb]
 \plotone{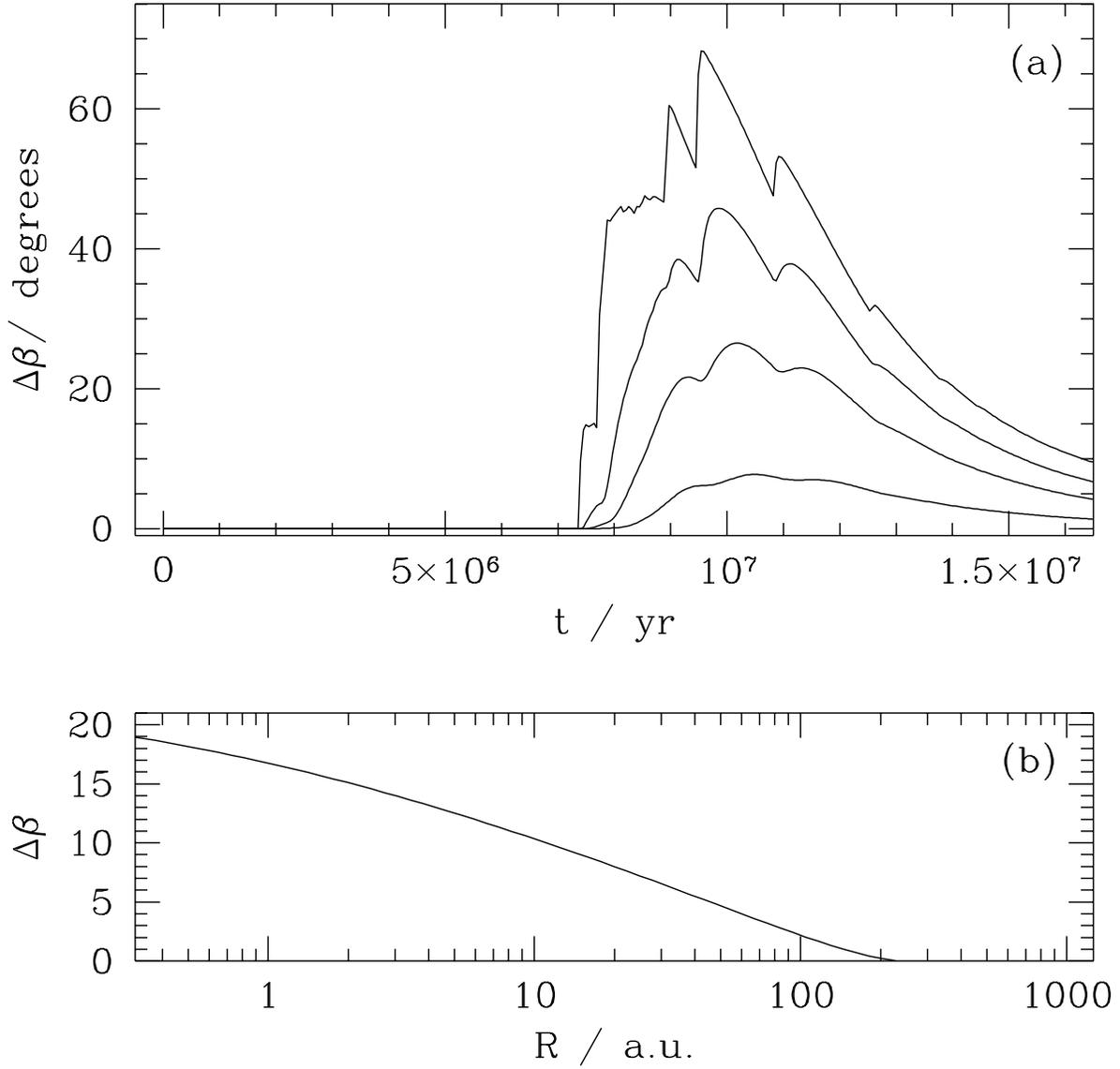}
 \caption{(a) The angle of tilt $\Delta \beta$ of annuli in the disk, 
 relative to the outermost annulus at 250 a.u. From top downwards, the annuli
 are at 3, 10, 30 and 100 a.u. (b) The tilt of the disk as a 
 function of radius at $t \approx 1.6 \times 10^7$ years, well 
 after radiative excitation of the warp has ceased.}
\end{figure} 

\end{document}